\documentclass[fleqn,10pt]{wlscirep}
\usepackage{amsmath}
\usepackage{lineno}
\usepackage{txfonts}
\usepackage{color}
\title{Controlling invasive ant species: a theoretical strategy for efficient monitoring in the early stage of invasion}

\author[1,*]{Shumpei Ujiyama}
\author[2]{Kazuki Tsuji}
\affil[1]{Innovation Science Course, School of Environment and Society, Tokyo Institute of Technology, 2-12-1 Ookayama, Meguro-ku, Tokyo, 152-8550, Japan}
\affil[2]{Department of Subtropical Agro-Environmental Sciences, University of the Ryukyus, Senbaru 1, Nishihara, Okinawa, 903-0213, Japan}

\affil[*]{ujiyama.s.aa@m.titech.ac.jp}

\begin{abstract}
Invasion by the red imported fire ant, \textit{Solenopsis invicta} Buren, has destructive effects on native biodiversity, agriculture, and public health. This ant's aggressive foraging behaviour and high reproductive capability have enabled its establishment of wild populations in most regions into which it has been imported. An important aspect of eradication is thorough nest monitoring and destruction during early invasion to prevent range expansion. The question is: How intense must monitoring be on temporal and spatial scales to eradicate the fire ant? Assuming that the ant was introduced into a region and that monitoring was conducted immediately after nest detection in an effort to detect all other potentially established nests, we developed a mathematical model to investigate detection rates. Setting the monitoring limit to three years, the detection rate was maximized when monitoring was conducted shifting bait trap locations and setting them at intervals of 30 m for each monitoring. Monitoring should be conducted in a radius of at least 4 km around the source nest, or wider---depending on how late a nest is found. For ease of application, we also derived equations for finding the minimum bait interval required in an arbitrary ant species for thorough monitoring.
\end{abstract}

\DeclareMathOperator*{\undr}{\iint}

\begin{document}

\flushbottom
\maketitle
\thispagestyle{empty}

%INTRODUCTION
%\section*{Introduction}

\noindent Invasive ants may have destructive effects on native biodiversity, agriculture and public health \cite{Porter&Savignano:1990,Allen_etal:1994,Gotelli&Arnett:2000,Wojcik_etal:2001,Holway_etal:2002,Morrison:2002,Allen_etal:2004}. One of the most harmful of these ants is the red imported fire ant, \textit{Solenopsis invicta} Buren, which is highly aggressive and can sting humans; increasing numbers of people are suffering the sequelae of these stings \cite{Rhoades_etal:1989,Kemp_etal:2000}. Clinical reactions to fire ant stings range from mild discomfort to life-threatening anaphylaxis \cite{Hoffman:1995,LaShell_etal:2010}; surveys reported that 0.6 \% to 6 \% of individuals who are stung have anaphylactic reactions \cite{deShazo_etal:1999}. Public health is not the only concern: fire ants can interfere with mechanical and electrical infrastructure and cause short circuits, resulting in malfunction of the equipment like traffic signals \cite{NEWYORKTIMES:1990,MacKay_etal:1992}. The economic impact of fire ant infestations is growing along with their range expansion, with the estimated costs of control, medical treatment, and damage to property in the United States exceeding \$1 billion annually by the beginning of the twenty-first century \cite{Pimentel_etal:2005}.

The fire ant was inadvertently introduced into the United States between 80 and 90 years ago \cite{Buren:1972,Buren_etal:1974}, and it spread rapidly throughout the southern part of the country. More recently, it has been introduced to other regions of the world, including the Caribbean, Australia, New Zealand, China, Taiwan, Japan and South Korea. With its aggressive foraging behaviour and high reproductive capability \cite{Tschinkel:1998,Allen_etal:2004}, the fire ant has established wild populations in all regions into which it has been imported, with the exception of New Zealand, Japan, and South Korea. Since 2001, the fire ant has been imported into New Zealand at least twice \cite{Bissmire:2006,Christian:2009}, but has been successfully eradicated. More recently, in 2017 the fire ant was discovered in ports in Japan and South Korea \cite{Lee:2017}, and the nests were immediately destroyed. It was detected in containers and within container yards; the fact that no other wild colonies have been found in these two countries indicates that these countries are not invaded yet \cite{Richardson_etal:2000}. However, the opportunity for species introduction is increasing because of the recent globalisation of the economy \cite{Bertelsmeier_etal:2017},  it would therefore not be surprising to find wild colonies in these and other uninvaded countries today.

One important aspect of ant eradication is thorough monitoring and destruction of nests in the early stage of invasion to prevent range expansion \cite{Simberloff:2003,Lodge_etal:2006}. However, once an index case is found, the following questions need to be answered before planning monitoring: (i) Where should we monitor? (ii) How intense should monitoring be on a spatio-temporal scale to enable the detection of all nests that may have simultaneously been established? And (iii) how can we minimise the costs of monitoring? The first question may be answered by potential distribution models \cite{Korzukhin_etal:2001,Ward:2007} and habitat-suitability models \cite{Alston-Knox_etal:2017,Cacho_etal:2011}. The third question may be answered partly by the answer to the first question, and by referring to the recent advances in surveillance methodologies and management protocols \cite{McNicol:2006,Hoffmann_etal:2010}. However, studies to answer the second question are scarce. The accumulation of knowledge on ant eradications worldwide \cite{Hoffmann_etal:2016} would be helpful, however, tools that help decision makers to estimate the probability of detection and costs are lacking. To answer the question we need a spatially explicit model that incorporates the fire ant's existence probability distributions, the radius of the monitoring range, and the spatio-temporal frequency of monitoring.

Here, we develop a mathematical model that incorporates these parameters and investigate the efficiency of monitoring strategies for detecting introduced fire ants in the early stage of invasion. Our focus is to investigate the detection rate with variations in the monitoring area and the spatio-temporal intensity of monitoring---not the efficiency of the specific traps utilized. 

%METHOD
\section*{Material and Methods}
\subsection*{Assumptions}
We assume that the red imported fire ant was introduced to an uninvaded region and settled in the wild \cite{Richardson_etal:2000}. The nest was found some time after the nest started producing alate queens, and monitoring started then. We assume that monitoring is conducted with highly attractive bait traps, such as soybean-oil-absorbed corn grits and hotdogs \cite{Lofgren_etal:1975,Williams_etal:2001}. We also assume that, with such a bait trap, a fire ant nest is detected at 100\% probability when the bait is placed in the foraging territory around the fire ant nest, except when the nest is at the incipient stage (See Supplementary Information (SI) section 1 for further explanation.). For ease of application, bait traps are set in lattice patterns. We set the origin of the coordinate axes to the location where the first-generation nest (here called the ``source nest'') was found. Assume that the existence probability distribution of a next-generation nest at location $(x,y)$ varies with the dispersal capability of an alate queen, i.e. the location of a nest is determined solely at the time of independent founding after the nuptial flight, and nest relocation \cite{Hays_etal:1982} is thereafter not considered. The fire ant has two social forms: monogyne and polygyne. A long-distance nuptial flight, followed by independent colony founding, takes place only when the colony is in the monogynous form \cite{Tschinkel:2006}. Here, we assume the monogynous form, because in this form the colony expands its range faster \cite{Markin_etal:1971,Tschinkel:1998} and is thus more difficult to detect in monitoring than the polygynous form \cite{Henshaw_etal:2005}.

\subsection*{Cases considered: optimistic and pessimistic}
\begin{figure}[ht]
\centering
\includegraphics[width=\linewidth]{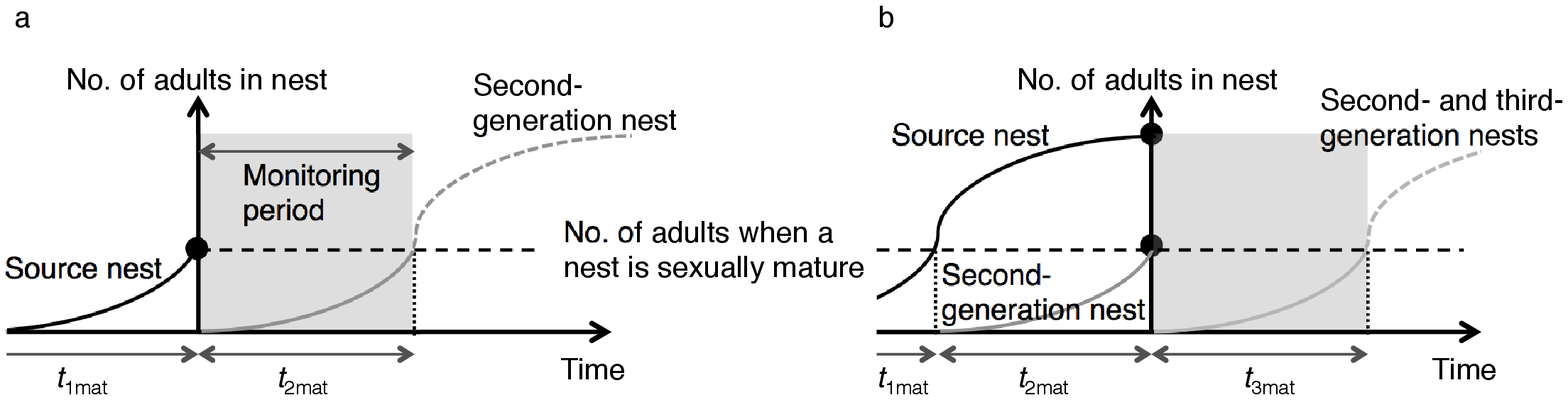}
\caption{\label{fg1}}
\end{figure}

We will consider two cases: the optimistic case, in which the source nest is found instantly after it starts producing queens, and second-generation alate queens have dispersed only for a short period (Figure \ref{fg1}a); and the pessimistic case, in which detection of the source and the second-generation nests is delayed, such that the second-generation nests have started producing queens and third-generation alate queens have dispersed for a short period (Figure \ref{fg1}b). The optimistic case corresponds to the situation in which the source nest has remained undetected for a while and the nest has matured to form a mound that someone has discovered by chance. The pessimistic case corresponds to the situation in which the source nest and the second-generation nests have been undetected for a while, and the second-generation nests have matured to form mounds, one of which has been noticed by someone. The source nest is assumed to be found approximately at the same time as the second-generation nest, because it is large and easy to detect once the searchers have been alerted. Note that the monitoring period should be no more than two to three years in both cases, because the next-generation nests will become sexually mature and produce alate queens in two to three years \cite{Markin_etal:1973}, depending on the food density and climatic conditions \cite{Tschinkel:1988}. In other words, monitoring has to be intense enough to be able to detect all of the dispersed nests within two to three years.

\subsection*{Dispersal of alate queens}
Surveys show that more than 99\% of alate queens disperse within 2 km of the source nest \cite{Markin_etal:1971}, and a simple energetic model suggests that flight capability of alate queens is limited to less than 5 km in the absence of wind \cite{Vogt_etal:2000}. Our focus is to investigate the detection rate with variations in the monitoring area and the spatio-temporal intensity of monitoring---not the dispersal kernel of fire ants in the presence of wind. Therefore, in this paper we assume the absence of wind such that inseminated queens disperse less than 5 km---that is, second-generation nests are distributed within 5 km of the source nest, and third-generation nests (in the pessimistic case) are distributed within 10 km of the source nest (see Supplementary Fig. S4). Because monitoring for 5 or 10 km around the source nest could be a costly task, we will consider cases in which the monitoring area does not fully cover the area in which the fire ants are distributed. As the detection rate is expected to increase if the bait locations are shifted for repeated monitoring (see later), we shall consider this case too.

\subsection*{Definition of the detection rate}
Given the above assumptions, the \textit{detection rate} $D(t)$ is defined as follows:
\begin{equation}
	D(t)=A_\mathrm{m} \times O(t), \label{eq1}
\end{equation}
where $A_\mathrm{m} (0 \le A_\mathrm{m} \le 1)$ and $O(t) (0 \le O(t) \le 1)$ denote the thoroughness and effectiveness of monitoring, respectively, (see SI section 3 for derivation) and time $(t)=0$ when monitoring starts.

$A_\mathrm{m}$ is the ratio of the monitoring area to the entire area where second- or third- generation nests (or both) may exist.
\begin{equation}
	A_\mathrm{m}=\undr_{-r_\mathrm{m} \le x,y \le r_\mathrm{m}}P(x,y)\mathrm{d}x\mathrm{d}y, \label{eq2}
\end{equation}
where $P(x,y)$ is the existence probability distribution of second- and third-generation nests, and $r_\mathrm{m}$ is the \textit{monitoring range} in the direction of the x- and y- axes from the origin; that is, the monitoring area is the square of the monitoring range. $P(x,y)$ denotes the probability that a second- or third- generation queen establishes a nest at location $(x,y)$ (Figure \ref{fg2}).

$O(t)$ is the \textit{observable ratio}, that is, the ratio of the \textit{detectable area} to the area of a square surrounded by four baits (see Supplementary Fig. S5). A detectable area is an area inside which fire ants may be caught and detected should a nest exist there. It is the area inside a circle, the centre of which is positioned on a bait (as in Supplementary Fig. S6); the radius of the detectable area, $r(t)$ is $r(t)=r_\mathrm{c}(t)+r_\mathrm{s}(t)$, where $r_\mathrm{c}(t)$ is the \textit{radius of a nest} (or nest mound) and  $r_\mathrm{s}(t)$ is the \textit{radius of a foraging territory} (see SI section 3 for a full explanation). When the bait location is not shifted, $O(t)$ is given as follows. (See SI section 4 for the case in which bait location is shifted.)
\begin{equation}
	O(t)=	\begin{cases}
	\piup \left( \frac{r(t)}{l_\mathrm{b}} \right)^2 & 0 \le r(t) < l_\mathrm{b}/2 \\
	\piup \left( \frac{r(t)}{l_\mathrm{b}} \right)^2 - \frac{4r(t)^2 \left( \arccos{\frac{l_\mathrm{b}}{2r(t)}} - \frac{l_\mathrm{b}}{2r(t)} \sin{\left( \arccos{\frac{l_\mathrm{b}}{2r(t)}} \right)} \right)}{l_\mathrm{b}^2}  & l_\mathrm{b}/2 \le r(t) < l_\mathrm{b}/\sqrt{2} \\
	1.0 & r(t) \ge l_\mathrm{b}/\sqrt{2}
	\end{cases}
	\label{eq3}
\end{equation}
Here, $l_\mathrm{b}$ is the \textit{spatial interval of baits} (or ``bait interval''). The second term in the middle formula represents the overlap of the detectable area. $O(t)$ increases if $r(t)$ increases and decreases if $l_\mathrm{b}$ increases. Thus, the second term in equation (\ref{eq1}) denotes the radius- and bait-interval-dependent probability that a fire ant nest will be detected at time $t$. $r_\mathrm{c}(t)$ and $r_\mathrm{s}(t)$ are defined as functions of the number of adults in a nest, $S(t)$, as follows:
\begin{equation}
	r_\mathrm{c}(t)=\sqrt{\frac{S(t)}{\piup d}}, \label{eq4}
\end{equation}
\begin{equation}
	r_\mathrm{s}(t)=\sqrt{\frac{\beta \cdot S(t-t_\mathrm{s})}{\piup}}, \label{eq5}
\end{equation}
\begin{equation}
	S(t)=\frac{220000}{1+83 \mathrm{e}^{-1.26 t}}. \label{eq6}
\end{equation}
Here, $\beta$ is a ratio to convert the number of adults in a nest to territorial area, $t_\mathrm{s}$ is the age of a nest when adults start searching for resource, and $S(t)$ is obtained empirically \cite{Tschinkel:1988}. See Table \ref{tb1} for the values of parameters and constants. $d$ is deduced from equation (\ref{eq4}) and the assumption that the radius of a fully-grown ($S(t)=22,000$) nest is 15 to 20 m (including underground tunnels), depending on the soil conditions. $\beta$ is deduced from equation (\ref{eq5}) and the results of a survey that the territorial area is approximately 1/1000 times the number of adults in a nest \cite{Tschinkel_etal:1995}. Worker ants start searching for resources two to four weeks (roughly $1/17$ of a year) after the nest is established.

Monitoring a small area via densely set baits means low thoroughness ($A_\mathrm{m}$) and high effectiveness ($O(t)$), which would result in low $D(t)$. Similarly, monitoring a large area via sparsely set baits means high thoroughness and low effectiveness, which would also result in low $D(t)$.

\subsection*{Data Availability}
All data generated and analysed during this study are included in this article (and its Supplementary Information file). More details are available from the corresponding author on reasonable request.

%RESULTS
\section*{Results}

\subsection*{Three-year detection rates when bait location is not shifted}
First, we will assume that the locations of baits are not shifted and monitoring is conducted once at $t=3$ years. $O(t)$ increases logistically until its value reaches one (Figure \ref{fg3}a), because it is dependent on the radius of the nest and the radius of the territory (Figure \ref{fg3}b), and both radiuses are dependent on the number of adults in the nest, which also increases logistically (Figure \ref{fg3}c).

We plotted the detection rate at $t=3$ years rounded to two decimal places (the \textit{three-year detection rate}) as a function of monitoring range around the source nest ($r_\mathrm{m}$, in metres) and the bait interval (Figure \ref{fg4}). $D(t)=1$ implies that if the fire ants have dispersed from the source nest, then they will all be detected.

\begin{figure}[ht]
\centering
\includegraphics[width=\linewidth]{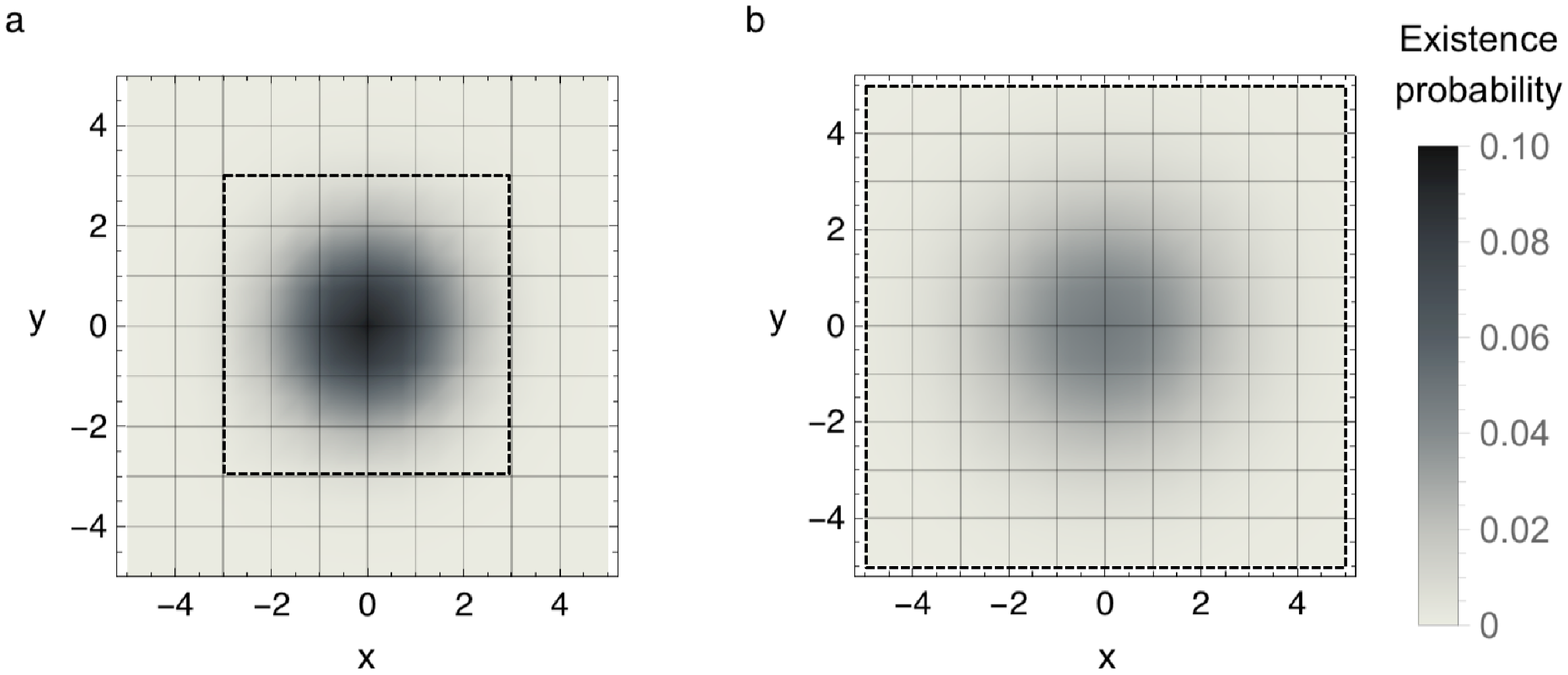}
\caption{\label{fg2}}
\end{figure}

\begin{figure}[ht]
\centering
\includegraphics[width=\linewidth]{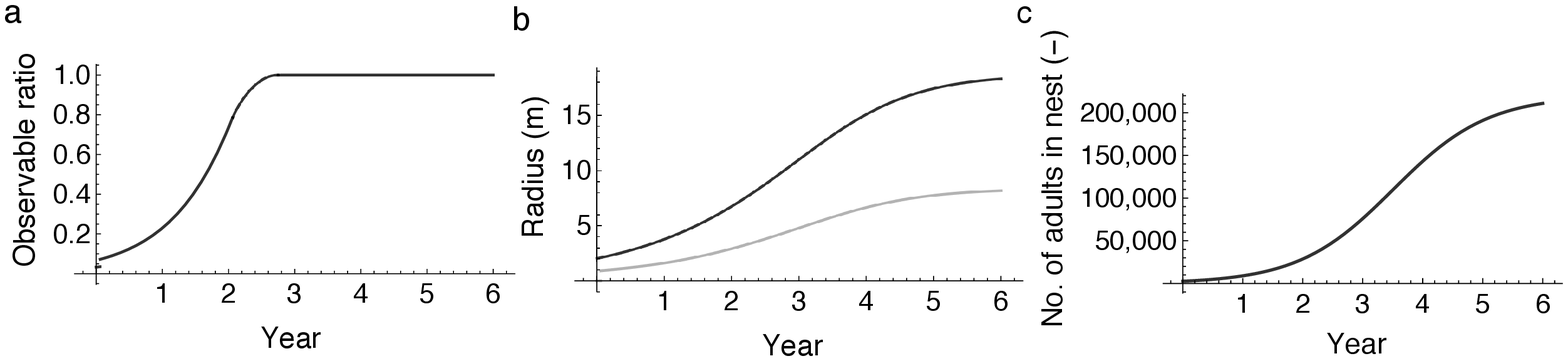}
\caption{\label{fg3}}
\end{figure}

\begin{figure}[ht]
\centering
\includegraphics[width=\linewidth]{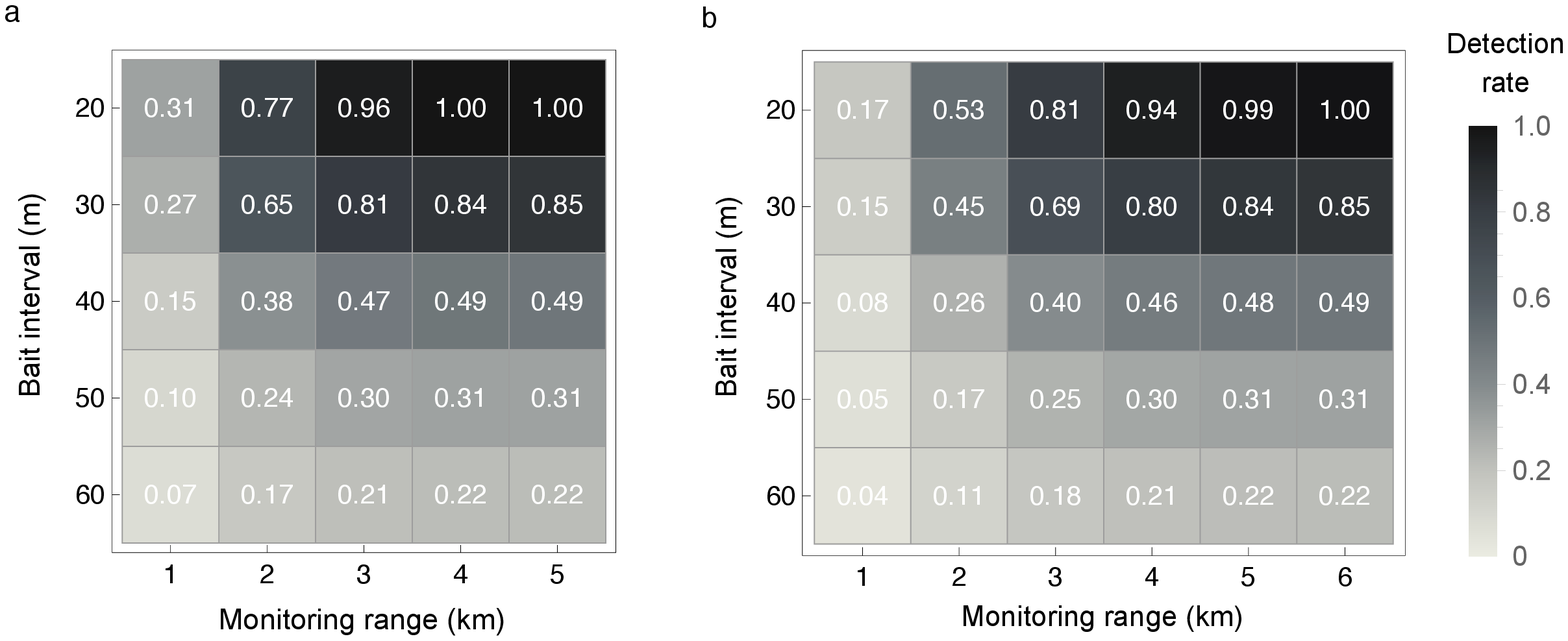}
\caption{\label{fg4}}
\end{figure}

Expanding the monitoring range increases the detection rate, especially up until a range of 3 km. The rate of increase in detection rate gradually slows, because the existence probability of fire ants become low over 3 km in the optimistic case (Figure \ref{fg2}a) and 5 km in the pessimistic case (Figure \ref{fg2}b). The detection rate is approximately one when 4 km around the nest is monitored in the optimistic case or 6 km around the nest is monitored in the pessimistic case. Detection rates when the monitoring range exceeded 6 km were omitted in the pessimistic case (Figure \ref{fg4}b) because they were approximately one.

When the entire area where fire ants may exist is monitored, it is only when the bait interval $l_\mathrm{b}$ is 20 m that the detection rate becomes one within three years (Figure \ref{fg4}). This finding implies that the bait interval has to be small to completely detect fire ants. Furthermore, when the monitoring range is reduced to reduce the cost, the three-year detection rate is as low as 31\% (optimistic case) or 17\% (pessimistic case) with the smallest bait interval and smallest monitoring range (Figure \ref{fg4}, $l_\mathrm{b}=20$ m, monitoring 1 km around the source nest).

Comparison of the results of the optimistic and pessimistic cases shows that, even with investment in the same monitoring cost (i.e. use of the same monitoring range and bait interval), the three-year detection rate is 24\% higher in the optimistic case than in the pessimistic one (Figure \ref{fg4}, $l_\mathrm{b}=20$ m, monitoring for 2 km around the source nest). This suggests that early detection of source and second-generation nests is vital.

\subsection*{Three-year detection rates when bait location is shifted}
\begin{figure}[ht]
\centering
\includegraphics[width=\linewidth]{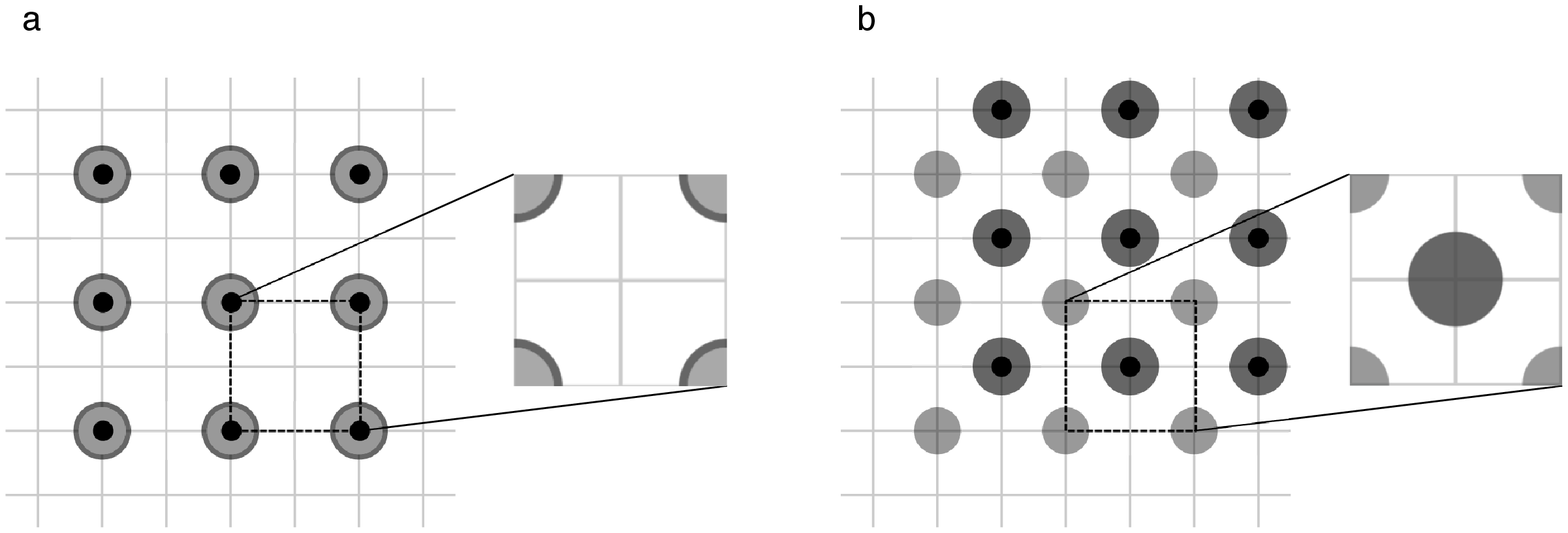}
\caption{\label{fg5}}
\end{figure}

Now we will assume that the locations of baits are shifted for the repeated monitoring: baits are located at $(x_i^t,y_j^t)=(i \times l_\mathrm{b},j \times l_\mathrm{b})$ $(i,j=0,\pm 1,\pm 2,...,\pm 2 \times r_\mathrm{m}/l_\mathrm{b})$ during the first monitoring at time $t=2.5$, and at $(x_i^{t+t_{\mathrm{int}}},y_j^{t+t_{\mathrm{int}}})=(x_i^t+l_\mathrm{b}/2,y_j^t+l_\mathrm{b}/2)$ in the subsequent monitoring conducted at time interval $t_{\mathrm{int}}=0.5$ after the first (see Supplementary Fig. S6). In other words, the bait locations are shifted to half the bait interval in the directions of the x- and y-axes. When the locations of baits are shifted, the area detectable through repeated monitoring expands (Figure \ref{fg5}), thus improving the detection rate. Remarkably, the bait interval can be increased 1.5 times, from 20 m to 30 m, by revising the bait location, while maintaining the detection rate at one (Compare Fig. \ref{fg4}a with Fig. \ref{fg6}a, and Fig. \ref{fg4}b with Fig. \ref{fg6}b). This implies that the monitoring cost could be cut roughly by half---the inverse of square of 1.5 to be precise, when shifting the location of baits.

We summarised the cost of monitoring, number of bait traps required, when bait location is not shifted and shifted (Table \ref{tb2}). The least-expensive monitoring strategy in the optimistic case is to shift the location of baits and monitor 4 km around the source nest with bait-intervals of 30 m, and 6 km around the source nest in the pessimistic case. Although there are two monitoring when the location of baits are shifted, the cost of monitoring is less than the case of which the bait location are not shifted, because the cost decreases the inverse of square of 1.5 times by revising the bait location.

%DISCUSSION
\section*{Discussion}

We used mathematical modelling to study the efficiency of monitoring strategies to detect and eradicate invasive ants in the early stage of their invasion. Setting the time limit to three years, the most efficient monitoring strategy is to conduct monitoring shifting the locations of traps and setting them at intervals of 30 m in each monitoring  (Table \ref{tb2}).

\begin{figure}[ht]
\centering
\includegraphics[width=\linewidth]{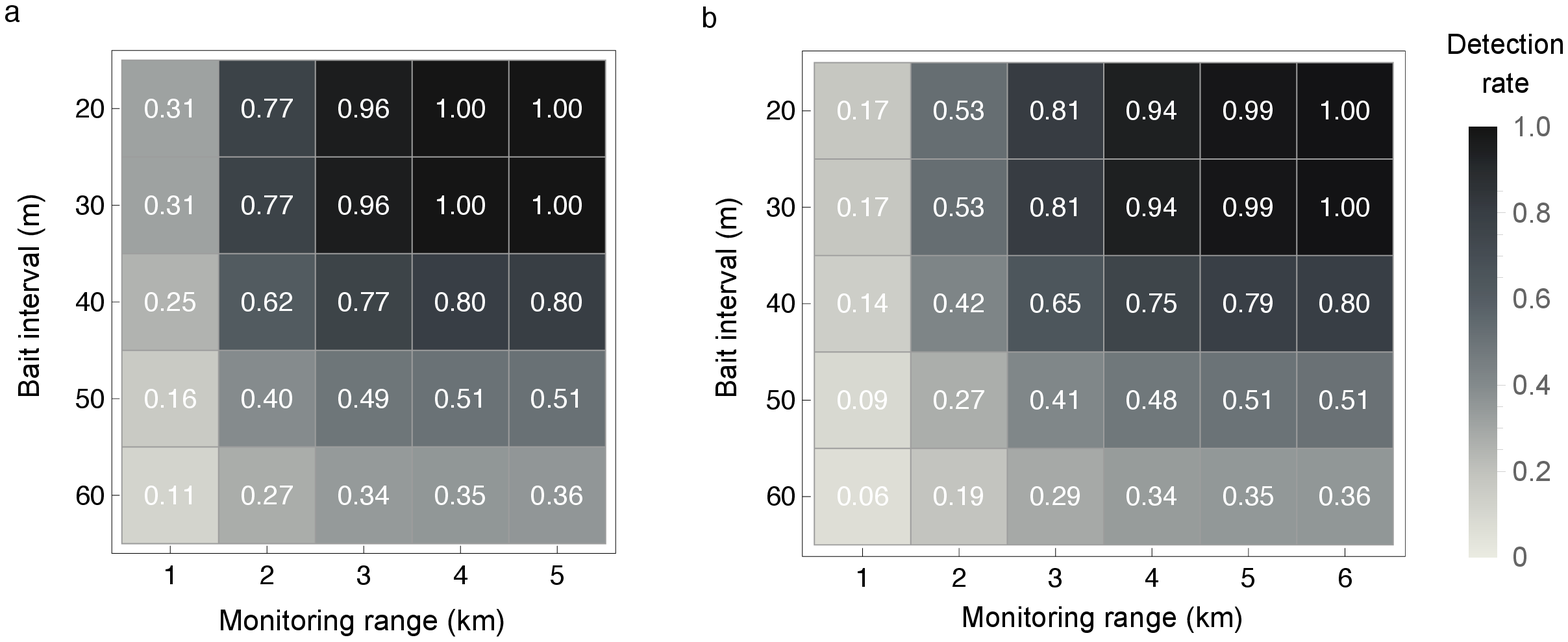}
\caption{\label{fg6}}
\end{figure}

The preferred monitoring range depends on how early or late the source nest is found. In the optimistic case (i.e. the source nest is found immediately after alate queen production starts), the preferred monitoring range is a radius of 4 km around the source nest. In the pessimistic case (i.e. the second-generation nests are found three years after alate queen production starts, so that the third-generation nests have become established), the preferred monitoring range is a radius of 6 km. Reducing the monitoring range to 3 km or less in the optimistic case and 5 km or less in the pessimistic case may allow fire ants to spread and thus should be avoided (Figure \ref{fg4} and Figure \ref{fg6}).

Early detection of first- and second-generation nests is vital, because the three-year detection rates differ substantially between the optimistic and pessimistic case when the monitoring cost is the same. (Compare Fig. \ref{fg4}a with Fig. \ref{fg6}a, and Fig. \ref{fg4}b with Fig. \ref{fg6}b.) We ignored the relocation of nests here. However, if relocation is considered, then shifting of trap locations will not be effective; the results for the case in which trap locations are not shifted should be used to assess the efficiency of monitoring in the case of nest relocation (Figure \ref{fg4}).

We assumed here that bait traps were used, but the type of trap used will vary with the object species. Actual detection rates are subject to the efficacy of individual bait traps; use of the appropriate trap therefore becomes important in applying these results to eradication programs. The efficacy of traps for certain species was not our target here, but it has been explored elsewhere \cite{Greenslade&Greenslade:1971,Romero&Jaffe:1989}. Recall that one of the important conclusions of this study is the determination of the recommended monitoring range, that is a radius of 4 or 6 km from the source nest. If the traps used are not as highly attractive as is assumed in our model \cite{Stringer_etal:2011}, or the trap efficacy decreases depending on distance from the nest entrance \cite{Stringer_etal:2011}, then, within this 4 or 6 km radius of the monitoring range, monitoring should be conducted not solely by traps but by a combination of visual surveillance\cite{McNicol:2006}, high-tech surveillance (such as drones with special cameras adopting image recognition technologies), and fire-ant-sniffing dogs \cite{Lin_etal:2011,Ward_etal:2016} to achieve thorough monitoring. We showed that repeated monitoring increases the detection rate. This also holds true in the case of high detection error ---further repetition may be recommended to increase detection rate--- however this is a matter of degree. In practice, if a large nest mound was discovered during the repetition, this implies that alate queens may have dispersed from that nest, thus further monitoring should be conducted setting the newly-discovered nest as the centre of monitoring range.

Note that the average dispersal distance of fire ants depends on the air temperature: the hotter the ambient temperature, the farther the dispersal. The monitoring area therefore would need to be expanded in tropical and subtropical regions. Colony growth rate also depends on temperature. We set the time limit of monitoring to three years, assuming the climate of mainland Japan, because independently founded colonies start to produce alate queens within three years under climatic conditions that are close to those at the northern end of the fire ants' potential distribution \cite{Morrison_etal:2004}. The monitoring time limit should therefore be shorter in warmer regions such as Okinawa and Taiwan.

Detection of the source nest is crucial, because setting the centre of the monitoring range on a second- or third- generation nest would reduce the thoroughness of detection (see SI section 5). To avoid this error, the age of the nest should be estimated before the nest is destroyed. Two parameters might be helpful in estimating nest age: one is the number of adults in the nest, and the other is the age of the queen ant. Estimating nest age from the number of adults in the nest requires the estimation of a number of parameters such as intrinsic growth rate and carrying capacity; therefore, the utilization of queen age may be better (see reference \cite{Tschinkel:1987}). However, collection of the queen in a monogynous field colony is also a difficult task. If multiple nests ---all similar in size--- were found, and estimating the nest age turned out to be difficult, then monitoring should be conducted within 6 km radius from each of the nests.

To conduct monitoring efficiently, it is vital to know accurately and precisely the dispersal kernel of fire ants. We assumed a Gaussian distribution type of kernel, but if the actual kernel is more leptokurtic \cite{Kot:1996,Clobert_etal:2012}, or the average dispersal distance is shorter \cite{Markin_etal:1971}, or the dispersal dynamics is subject to interspecific interactions \cite{Hasting_etal:2005}---or any combination of these factors--- then the detection rate increases more rapidly with smaller monitoring range.

The model presented here is applicable not only to fire ants, but also to a wide range of organisms that (i) establish colonies and (ii) have a dispersal kernel and a territorial area that can be defined. We calculated the detection rate as a function of bait interval and other parameters, but it would be convenient to have an equation to derive the minimum $l_\mathrm{b}$ (the spatial interval of traps such as baits) required for the detection rate to be one until an arbitrary time $t$. The equations below give the minimum trap intervals when we assume that the entire area in which the object species may exist is monitored. Equation (\ref{eq7}) covers the case in which trap location is not shifted or the object species frequently relocates its nests, whereas equation (\ref{eq8}) covers the case in which trap location is shifted (see SI section 6 for derivation).
\begin{equation}
	l_\mathrm{b}=\sqrt{2}r(t), \\ \label{eq7}
\end{equation}
\begin{equation}
	l_{\mathrm{b,shift}}=r(t)+\sqrt{2 \cdot r(t+t_{\mathrm{int}})^2-r(t)^2}. \label{eq8}
\end{equation}
See Table \ref{tb1} for the parameter list. Note that $t+t_{\mathrm{int}}\le3$ if the object species is \textit{S. invicta}, in order to avoid the production of next-generation queens. Substitute $d = 200$, $t_\mathrm{s} = 1/17$, $\beta = 1/1000$, and assume $t=3$ for equation (\ref{eq7}), and $t=2.5$ and $t_{\mathrm{int}}=0.5$ for equation (\ref{eq8}). Then, $l_\mathrm{b}=22.3$ and $l_{\mathrm{b,shift}}=31.0$. These values are confirmed as reasonable from our findings that the detection rate was one up to $l_\mathrm{b}=20$ in Fig. \ref{fg4} and up to 30 m in Fig. \ref{fg6}. In applying these equations to species other than \textit{S. invicta}, the species-specific functions of $r(t)$ and $S(t)$ have to be empirically determined. In practice, growth of the colony's territorial area may often be easier to monitor than that of colony size, because estimating accurately the colony size requires excavation of the whole nest in which the radius is 15 m or larger depending on soil condition. Note that the territorial area may change on a daily basis \cite{Showler_etal:1990}, thus it is important to collect data for several days and use the average value.

%REFERENCE
\bibliography{reference}

%ACKNOWLEDGEMENT
\section*{Acknowledgements}
None of the authors had personal or financial conflicts of interest.

%CONTRIBUTION
\section*{Author Contributions}
S.U. performed the research with input from K.T., and S.U and K.T. wrote and reviewed the manuscript.

%ADDITIONAL INFORMATION
\section*{Additional Information}
\textbf{Supplementary information} accompanies this paper at doi: (link to be determined)\\
\textbf{Competing financial interests:} The authors declare no competing financial interests.
%To include, in this order: \textbf{Accession codes} (where applicable); \textbf{Competing financial interests} (mandatory statement). 

%FIGURE LEGENDS
\newpage
\noindent Figure Legends.\\\\
\noindent \textbf{Figure 1. Monitoring periods in the optimistic and pessimistic cases.} Horizontal and vertical axes denote time and number of adults in the nest, respectively. Black curve, transition in the number of adults in the source nest; grey curve, transition in the number of adults in the second-generation nest; light grey curve, transition in the number of adults in the third-generation nest. $t_{\mathrm{1mat}}$, $t_{\mathrm{2mat}}$, and $t_{\mathrm{3mat}}$ are the times when the source, second-generation, and third-generation nests, respectively, become sexually mature. Black dots indicate times of detection and destruction of focal nests. Lightly shaded region denotes the monitoring period. (a) Optimistic case. (b) Pessimistic case.\\\\

\noindent \textbf{Figure 2. Existence probability distribution $P(x,y)$.} Assuming that an alate queen has dispersed from the origin (0,0), the maps show the two-dimensional distributions of the existence probability of a next-generation nest at coordinates $(x,y)$. The square area surrounded by the broken line denotes the monitoring range, which is a radius of $r_\mathrm{m}$ around the origin. These figures give examples when $r_\mathrm{m}=3$ km in the optimistic case and $r_\mathrm{m}=5$ km in the pessimistic case. (a) Distribution of second-generation nests in the optimistic case. (b) Distribution of third-generation nests in the pessimistic case (see SI section 2 for derivation).\\\\

\noindent \textbf{Figure 3. Characteristic traits of nests.} (a) Number of adults in a nest $S(t)$. (b) Radius of nest $r_\mathrm{c}(t)$ (black line) and radius of foraging territory $r_\mathrm{s}(t)$ (grey line). (c) Observable ratio $O(t)$.\\\\

\noindent \textbf{Figure 4. Three-year detection rates as a function of monitoring range and bait interval when baits are not shifted.} Each number at a given interval and range denotes the detection rate rounded to two decimal places. (a) Optimistic case. (b) Pessimistic case.\\\\

\noindent \textbf{Figure 5. Overall detectable area when monitoring is conducted twice.} Black dots denote baits. Detectable areas in the first (light grey) and second (dark grey) monitorings are superimposed to show the overall detectable areas. (a) Not shifting baits. (b) Shifting baits.\\\\

\noindent \textbf{Figure 6. Three-year detection rates when baits are shifted.} Each number at a given interval and range denotes the detection rate rounded to two decimal places. (a) Optimistic case. (b) Pessimistic case.\\\\

%TABLES
\newpage
\noindent Tables.\\\\
\begin{table}[ht]
\noindent \caption{\label{tb1}\textbf{Parameters and constants.}}
\centering
\begin{tabular}{llll}	\hline
  \rowcolor[gray]{0.8}	Parameters and Constants & Symbol & Value & Dimension	\\ \hline
  Parameters	\\
  ~~~~Monitoring range & $r_\mathrm{m}$ & - & (m)	\\
  ~~~~Radius of a nest or nest mound & $r_\mathrm{c}(t)$ & - & (m)	\\
  ~~~~Radius of foraging territory from the edge of a nest & $r_\mathrm{s}(t)$ & - & (m)	\\
  ~~~~Radius of detectable area, i.e. $r_\mathrm{c}(t)+r_\mathrm{s}(t)$ & $r(t)$ & - & (m)	\\
  ~~~~Number of adults in a nest & $S(t)$ & - & (-)	\\
  ~~~~Spatial interval of traps such as baits & $l_\mathrm{b}$ & - & (m)	\\
Constants	\\
  ~~~~Age of a nest when it is fully sexually mature & $t_{\mathrm{1mat}}, t_{\mathrm{2mat}}, t_{\mathrm{3mat}}$ & 3 & (years)	\\
  ~~~~Temporal interval of the first and second monitorings & $t_{\mathrm{int}}$ & 0.5 & (years)	\\
  ~~~~Number of adults per square meter in a nest & $d$ & 200 & (m$^{-2}$)	\\
  ~~~~Ratio to convert the number of adults in a nest to territorial area & $\beta$ & 1/1000 & (m$^2$)	\\
  ~~~~Age of a nest when adults start searching for resource & $t_\mathrm{s}$ & 1/17 & (years)	\\ \hline \\
\end{tabular}
\end{table}

\begin{table}[ht]
\noindent \caption{\label{tb2}\textbf{Number of traps required to conduct monitoring.} ``-" shows the parameter set for which the three-year detection rates would be below one. The values are calculated based on an assumption: there is one monitoring when the bait locations are not shifted; and there are two monitorings when the bait locations are shifted.}
\centering
\begin{tabular}{lccc}	\hline
  \\
   & \multicolumn{3}{c}{Monitoring Range (m)}	\\ \cline{2-4}
   \\
   Bait Interval (m) & 4000 & 5000 & 6000	\\ \hline
  Not shifting baits (one monitoring)	\\
  ~~~~Optimistic case\\
  ~~~~~~~~20 & 160801 & 251001 & 361201	\\
  ~~~~~~~~30 & - & - & -	\\
  ~~~~Pessimistic case\\
  ~~~~~~~~20 & - & - & 361201	\\
  ~~~~~~~~30 & - & - & -	\\
  Shifting baits (two monitorings)	\\
  ~~~~Optimistic case\\
  ~~~~~~~~20 & 321602 & 502002 & 722402	\\
  ~~~~~~~~30 & 143291 & 223558 & 321602	\\
  ~~~~Pessimistic case\\
  ~~~~~~~~20 & - & - & 722402	\\
  ~~~~~~~~30 & - & - & 321602	\\ \hline
\end{tabular}
\end{table}

%FIGURES
\newpage
\noindent Figures\\\\

\newpage

\newpage

\newpage

\newpage

\newpage

\end{document}